# Correlated Electronic Structure and Incipient Flat Bands of the Kagome Superconductor CsCr$_3$Sb$_5$

Yidian Li[1,*], Yi Liu[2,3,*], Xian Du[1,*], Siqi Wu[2], Wenxuan Zhao[1], Kaiyi Zhai[1], Yinqi Hu[1], Senyao Zhang[1], Houke Chen[4], Jieyi Liu[4], Yiheng Yang[4], Cheng Peng[4], Makoto Hashimoto[5], Donghui Lu[5], Zhongkai Liu[6,7], Yilin Wang[8,9], Yulin Chen[4,6,7,†], Guanghan Cao[2,†], and Lexian Yang[1,10,†]

[1]*State Key Laboratory of Low Dimensional Quantum Physics, Department of Physics, Tsinghua University, Beijing 100084, China.*

[2]*School of Physics, Zhejiang University, Hangzhou 310058, P. R. China*

[3]*Department of Applied Physics, Key Laboratory of Quantum Precision Measurement of Zhejiang Province, Zhejiang University of Technology, Hangzhou 310023, China*

[4]*Department of Physics, Clarendon Laboratory, University of Oxford, Parks Road, Oxford OX1 3PU, UK.*

[5]*Stanford Synchrotron Radiation Lightsource, SLAC National Accelerator Laboratory, Menlo Park, CA, USA*

[6]*School of Physical Science and Technology, ShanghaiTech University and CAS-Shanghai Science Research Center, Shanghai 201210, China.*

[7]*ShanghaiTech Laboratory for Topological Physics, Shanghai 200031, China.*

[8]*School of Emerging Technology, University of Science and Technology of China, Hefei 230026, China*

[9]*New Cornerstone Science Laboratory, University of Science and Technology of China, Hefei, 230026, China*

[10]*Collaborative Innovation Center of Quantum Matter, Beijing 100084, China.*

[*]These authors contributed equally to this work.

[†]e-mail: YLC: yulin.chen@physics.ox.ac.uk; GHC: ghcao@zju.edu.cn; LXY: lxyang@tsinghua.edu.cn.

Kagome materials exhibit many novel phenomena emerging from the interplay between lattice geometry, electronic structure, and topology. A prime example is the vanadium-based kagome materials $A$V$_3$Sb$_5$ ($A$ = K, Rb, and Cs) with superconductivity and unconventional charge-density wave (CDW). More interestingly, the substitution of vanadium by chromium further introduces magnetism and enhances the correlation effect in CsCr$_3$Sb$_5$ which likewise exhibits superconductivity under pressure and competing density-wave state. Here we systematically investigate the electronic structure of CsCr$_3$Sb$_5$ using high-resolution angle-resolved photoemission spectroscopy (APRES) and ab-initio calculations. Overall, the measured electronic structure agrees with the theoretical calculation. Remarkably, Cr 3$d$ orbitals exhibit incoherent electronic states and contribute to incipient flat bands close to the Fermi level. The electronic structure shows a minor change across the magnetic transition at 55 K, suggesting a weak interplay between the local magnetic moment and itinerant electrons. Furthermore, we reveal a drastic enhancement of the electron scattering rate across the magnetic transition, which is relevant to the semiconducting-like transport property of the system at high temperatures. Our results suggest that CsCr$_3$Sb$_5$ is a strongly correlated Hund's metal with incipient flat bands near the Fermi level, which provides an electronic basis for understanding its novel properties in comparison to the non-magnetic and weakly correlated $A$V$_3$Sb$_5$.

Kagome materials with corner-sharing triangular networks in their lattice structures exhibit rich and fascinating properties, such as frustrated lattice geometry for spin liquid states [1-3], non-trivial topological electronic structure for topological quantum phases [4-8], and destructive interference effects for fractional quantum Hall effect [9-11]. The characteristic electronic structure of a typical kagome lattice includes the flat band, the Dirac fermions at the $K$ point, and the van Hove singularity (vHS) at the $M$ point (Fig. 1a). If the Fermi level ($E_F$) is tuned to the flat band, novel magnetic states such as fractional quantum Hall effect and quantum anomalous Hall effect may emerge, while the vHS and the Dirac fermions at the $E_F$ can induce unconventional superconductivity, charge-density wave (CDW), and/or other novel properties [12-14]. Up to date, many different kagome materials with characteristic kagome electronic structure have been discovered and extensively studied [5, 8, 14-18]. However, the exploration of intrinsic kagome-related many-body ground states remains inadequate since the Dirac fermions at the $K$ point are usually fragile against spin-orbit coupling, while the flat band and vHS usually locate far away from $E_F$.

Among the abundant kagome materials, $A$V$_3$Sb$_5$ ($A$ = K, Rb, and Cs) have attracted great attention due to their intriguing emergent properties [19, 20], such as the interplay/competition between superconductivity and unconventional CDW [20-23], time- and/or rotational-symmetry broken phases [24-31], giant anomalous Hall effect related to the chirality of the charge order[27, 32], the observation of pair-density wave [33, 34], and putative loop current order[35-37]. In the exploration of $A$V$_3$Sb$_5$ materials, it has been a common consensus that the vHS near $E_F$ plays a key role in the CDW and possibly also in the superconductivity [38, 39], while the flat band was observed far away from $E_F$ [40].

Moreover, the non-magnetic and weakly correlated kagome physics of $A$V$_3$Sb$_5$ can be further enriched by introducing magnetism and strong electron correlations. By substituting vanadium with chromium, the sibling compound CsCr$_3$Sb$_5$ provides an ideal platform to investigate the impact of magnetism and electronic correlation effect in the $A$V$_3$Sb$_5$-type kagome systems [41]. Indeed, recent investigations reveal many novel properties of this new kagome material, including pressurized superconductivity at 6.4 K, frustrated altermagnetism, CDW-like order, and non-Fermi liquid behaviour in the normal state [41-43].

Compared to the non-magnetic and weakly correlated $A$V$_3$Sb$_5$, incipient flat bands (IFB, flat bands that extend in a small portion of Brillouin zone and locate away from $E_F$) [44] originated from correlated Cr $3d$ orbitals appear near $E_F$, which may dominate the correlated electronic properties of the system. These novel electronic characteristics, together with the prototypical kagome electronic structure of CsCr$_3$Sb$_5$, require systematic experimental studies of its electronic structure.

In this work, by performing high-resolution angle-resolved photoemission spectroscopy (ARPES) measurements, we systematically investigate the electronic structure of CsCr$_3$Sb$_5$ single crystals. Our experiments reveal weakly correlated Sb $5p$ states and incoherent Cr $3d$ states, in overall agreement with our density-functional-theory (DFT) and dynamical-mean-field-theory (DMFT) calculations. Compared to CsV$_3$Sb$_5$, the vHSs and Dirac fermions characterizing the kagome lattice locate far away from $E_F$. Interestingly, we observe a flat Cr $3d$ band at about 80 meV below $E_F$, confirming the existence of IFB close to $E_F$. At the lowest temperature (6 K), we do not observe clear band folding or splitting related to the magnetic transition, suggesting a weak coupling between the localized magnetic moment and itinerant conduction electrons. Moreover, the electron scattering rate as manifested by the spectral broadening is drastically enhanced above the magnetic phase transition, which can explain the insulating behavior of the system at high temperatures. Consistent with the theoretical calculations [45, 46], we confirm that CsCr$_3$Sb$_5$ is a strongly correlated Hund's metal with IFB. Our work provides a foundation for further exploration of the correlated kagome material with possible unconventional superconductivity.

CsCr$_3$Sb$_5$ crystallizes in a similar layered structure of $A$V$_3$Sb$_5$ with the space group of $P6/mmm$ (Fig. 1b) [41]. In each unit cell, the Cr atoms constitute a two-dimensional kagome lattice with Sb atoms occupying the center of the hexagons. It is mainly the Cr $3d$ orbitals that form the kagome electronic states near $E_F$, as schematically shown in Fig. 1c. Our susceptibility measurement suggests an antiferromagnetic transition at the Neel temperature $T_N = 55$ K, which significantly influences the resistivity (Figs. 1d, e), inducing metallic and semiconducting-like behaviours below and above $T_N$, respectively. Figures 1f, g compare the DFT-calculated electronic structure of CsCr$_3$Sb$_5$ and CsV$_3$Sb$_5$. In general, the two compounds show similar electronic structures characterizing the kagome lattice, including the vHS and the Dirac fermions. Compared to CsV$_3$Sb$_5$, where the vHSs are close to $E_F$, the vHS and Dirac fermions in CrCr$_3$Sb$_5$ locate further away from $E_F$, suggesting their irrelevance in the novel transport properties. Prominently, compared to CsV$_3$Sb$_5$, the band width of the dispersive bands just above $E_F$ is strongly suppressed, leaving flat bands slightly above $E_F$ in CsCr$_3$Sb$_5$ (Fig. 1f), which are believed to play a key role in the unconventional superconductivity [44, 47, 48].

Figure 2a shows the calculated three-dimensional Fermi surface (FS) of CsCr$_3$Sb$_5$. There is a nearly cylindrical electron pocket and two cylindrical hole pockets with similar volume at $k_\parallel = 0$ (the $\bar{\Gamma}$ point), together with a warped cylindrical electron pocket around $k_\parallel = 0.66$ Å$^{-1}$ (the $\bar{M}$ point). At certain $k_z$, there is no Fermi crossing along $\bar{\Gamma}\bar{K}\bar{M}$, leaving holes in the cylindrical hole pockets around $\bar{\Gamma}$ (see the yellow Fermi pockets in Fig. 2a). The experimental FS in the $k_x$-$k_z$ plane shows weak but resolvable $k_z$ dispersion of the bands around $k_\parallel = 0$ (the $\bar{\Gamma}$ point), confirming the quasi-two-dimensionality of the electronic structure (Fig. 2b). The FS in the $k_x$-$k_y$ plane shows a hexagonal structure (Fig. 2c). We observe a circular electron pocket around the $\bar{\Gamma}$ point, together with a spectral-weight patch around the $\bar{M}$ point. With increasing binding energy, the pocket around $\bar{\Gamma}$ shrinks and line-like features can be resolved, forming a kagome-shape structure in the momentum space.

Figures 2e-g show the experimental band dispersions along high symmetry directions in a large energy range. Since the electronic states near $E_F$ are mainly contributed by Cr 3$d$ orbitals, the data strongly depend on the light polarization. In general, the spectra can be roughly divided into three segments in energy: the dispersive bands near $E_F$, an M-shape feature around -1.5 eV, and weakly dispersive bands around 4 eV. These features are in overall consistent with the calculation in Fig. 1f, as confirmed by the comparison between the integrated energy distribution curve (EDC) and the calculated density of states in Fig. 2h.

Figure 3 shows the fine electronic structure along high-symmetry directions near $E_F$. The experiment shows an overall agreement with the calculation (Figs. 3a, b), as shown by the data overlaid with the calculation in the Supplementary Information. The energy bands originated from Sb $p$ orbitals are well captured by the calculation. Specifically, they form the vHS at Γ, which shows electron-like dispersion in the $k_x$-$k_y$ plane and hole-like dispersion along $k_z$, as schematically shown in Fig. 3c. For the Cr 3$d$ orbital bands, the spectra are generally much broader, with the spectral weight distribution consistent with the theoretically predicted Cr $d$ orbital bands. It is noteworthy that while non-trivial surface states related to the $Z_2$ topological electronic structure were observed in $A$V$_3$Sb$_5$, here we do not find clear evidence for the surface states.

Prominently, we reveal a flat band very close to $E_F$, as indicated by the red arrows in Fig. 3b. The flat band with band bottom at about 80 meV below $E_F$ can be better resolved from the data along $\overline{\Gamma K}$ and the EDC at $\overline{\Gamma}$ shown in Fig. 3d (also see the Supplementary Information). This flat band close to $E_F$ may dominate the transport properties of CsCr$_3$Sb$_5$ and contribute to the electronic specific-heat coefficient as large as 105 mJ/(K·mol) [41]. The observation of the flat band in ARPES experiment suggests that the DFT-calculated flat bands are pushed closer to $E_F$ and the experimental $E_F$ is slightly raised up. To understand this difference, we perform DFT+DMFT calculation, which can better capture the strong electronic correlations of Cr 3$d$ electrons. The calculated spectral function of the paramagnetic state at Coulomb interaction $U = 5$ eV and Hund's coupling $J_H = 0.88$ eV [45] is shown in Fig. 3e. Interestingly, the flat bands above $E_F$ in Fig. 3a are pushed closer to $E_F$ and strongly renormalized, leaving incipient flat band with prominent spectral weight mainly around $\overline{\Gamma}$, which can be observed if the experimental $E_F$ is slightly raised up. Moreover, the DFT+DMFT calculation suggests that the electronic correlation effect shows a minor impact on the Sb $p$ orbital bands but strongly broadens Cr $d$ orbital bands (Supplementary Information), leaving very incoherent spectra. We emphasize that the strong Hund's coupling is essential in understanding the strong renormalization of the effective mass and electron self-energy [45]. The agreement between our experiment and DMFT calculation confirms that CsCr$_3$Sb$_5$ is a strongly correlated Hund's metal with incipient flat bands.

Although CsCr$_3$Sb$_5$ shows a clear magnetic and resistive transition near 55 K, we did not observe clear signatures of band folding, splitting, or shift related to the transition at the lowest temperature (6 K). The temperature-dependent measurement suggests a minor change of the spectra between 15 K and 75 K (Figs. 4a, b), which indicates a weak interplay between the localized magnetic moment and the itinerant carriers, as commonly observed in many magnetic materials [49]. To better resolve the dispersion around $\bar{\Gamma}$ and track the temperature evolution of the band structure, we conducted laser-ARPES measurements with better resolutions. The spectrum in Fig. 4c shows a strong intensity near $E_F$. By fitting the momentum distribution curves (MDCs) to Lorentzian, we extract the band dispersion as shown by the black lines in Fig. 4c. The electron band tends to exhibit a flattened band top as approaching $E_F$, consistent with the calculation in Fig. 3a. This feature can be better resolved at high-temperature data after divided by the Fermi-Dirac distribution function (Supplementary Information).

Figure 4d further shows the temperature evolution of our laser-ARPES spectra. Indeed, no clear change of the spectra was observed across the magnetic transition at 55 K, as can be seen from the corresponding EDCs near $E_F$ in Fig. 4e. The metallic band structure, however, deviates from the observation of an insulating resistivity above 55 K. We notice that the spectra are suddenly broadened across the magnetic transition as shown by the MDCs near $E_F$ (Fig. 4f). By fitting the MDCs to Lorentzians (black lines in Fig. 4f), we extract the full-width-at-half-maximum (FWHM) of the MDCs and plot the results as a function of temperature in Fig. 4g. Apparently, with increasing temperature, the FWHM quickly increases above about 50 K, suggesting an enhanced electron scattering rate at high temperatures, which is likely due to the magnetic fluctuation effect. Therefore, based on the metallic band structure and drastically enhanced electron scattering rate, we argue that the resistive transition at 55 K may be of a magnetic origin. Further experimental and theoretical exploration are required to unravel the connection between the electronic structure and magnetic transition in this material.

The observation of the incipient flat band and the strong correlation nature of $CsCr_3Sb_5$ provides an electronic basis for understanding its novel properties. Firstly, the cooperation of Hund's coupling and the incipient flat bands significantly enhances the orbital-dependent electronic correlation effect [45, 46], similar to the physics in iron-based superconductors [50-52] and leading to the large effective mass of carriers.

Secondly, Hund's coupling induces the localization of the magnetic moments and strongly increases the imaginary part of the electron self-energy, giving rise to the incoherent Cr $3d$ states, which is further enhanced by the antiferromagnetic fluctuation effect, as observed in our temperature-dependent measurements. Finally, the incipient flat band is believed to be crucial for unconventional superconductivity [44, 47, 48]. It has been shown that the application of pressure can effectively tune the portions and width of the incipient flat bands [45], which may be important for the pressurized superconductivity in $CsCr_3Sb_5$ [41].

In summary, we systematically investigate the electronic structure of the correlated kagome material $CsCr_3Sb_5$. Our work reveals the characteristic electronic structure including the vHS and incipient flat bands. The magnetism of the system shows a minor impact on the electronic structure, but the electronic scattering is drastically reduced in the magnetic ordered state. In agreement with the DFT+DMFT calculation, our experimental results suggest that $CsCr_3Sb_5$ is a strongly correlated Hund's metal with incipient flat bands. Our work provides crucial information for understanding the novel properties of $CsCr_3Sb_5$ in comparison to the weakly-correlated and non-magnetic $AV_3Sb_5$ compounds.

**Methods**

**Sample growth and characterization**

Single crystals of $CsCr_3Sb_5$ were grown via a self-flux method. The details of the crystal growth can be seen in ref. 41. Hexagonal-shape crystalline flakes with typical size of $0.5 \times 0.5 \times 0.02$ mm$^3$ were harvested for the ARPES measurement. Before the ARPES experiment, crystals from the same batch were characterized by the x-ray diffraction, energy-dispersive x-ray spectroscopy, and measurements of electrical resistivity and magnetic susceptibility. The resistivity measurement was conducted using a standard four-terminal method. The magnetic measurements were performed on a Magnetic Property Measurement System (MPMS-3, Quantum Design).

**ARPES measurements**

Synchrotron-based ARPES measurements were conducted at beamline 5-2 in Stanford synchrotron lightsource (SSRL, proposal No. S-XV-ST-6370A). The samples were cleaved *in-situ* under ultra-high vacuum below $1 \times 10^{-10}$ mbar. Data were collected with a Scienta DA30L electron analyser. The total energy and angular resolutions were set to 15 meV and 0.2°, respectively.

Laser-ARPES measurements were conducted at Tsinghua University. The 7-eV laser was generated by frequency doubling in a KBBF crystal and focused on the sample by an optical lens with a beam spot of about 20 μm. CsCr$_3$Sb$_5$ single crystals were cleaved *in-situ* under ultra-high vacuum below $5 \times 10^{-11}$ mbar. Data were collected by a Scienta DA30L electron analyzer. The total energy and angular resolutions were set to 3 meV and 0.2°, respectively.

**Ab initio calculations**

First-principles band structure calculations were performed using Vienna ab initio simulation package (VASP) [53] with a plane wave basis. The exchange-correlation energy was considered under Perdew-Burke-Ernzerhof (PBE) type generalized gradient approximation (GGA) [54] with spin-orbit coupling included. The cutoff energy for the plane-wave basis was set to 500 eV. A Γ-centered k-point mesh of $12 \times 12 \times 8$ was adopted in the self-consistent calculations.

**DFT+DMFT calculations**

The fully charge self-consistent single-site DFT+DMFT calculations are performed using the DFT+eDMFT code [55, 56] based on the WIEN2K package [57]. We choose a large hybridization energy window from -10 to 10 eV. All the five Cr-3d orbitals are considered as correlated ones and rotationally-invariant local Coulomb interaction Hamiltonian is used, which is parameterized by on-site Hubbard $U$ and Hund's coupling $J_H$. We choose the "exact" double-counting scheme [58]. The continuous time quantum Monte Carlo [59] is used as impurity solver. The self-energy on real frequency is obtained by the analytical continuation method of maximum entropy.

**Data availability**

The data sets that support the findings of this study are available from the corresponding author upon request.

**Acknowledgments**

This work is funded by the National Key R&D Program of China (Grant No. 2022YFA1403200, 2022YFA1403100, and 2023YFA1406101) and the National Natural Science Foundation of China (No. 12274251). L.X.Y. acknowledges support from the Tsinghua University Initiative Scientific Research Program and the Fund of Science and Technology on Surface Physics and Chemistry Laboratory (No. XKFZ202102). Y.L.W. was supported by the National Natural Science Foundation of China (No. 12174365) and the New Cornerstone Science Foundation. Use of the Stanford Synchrotron Radiation Lightsource, SLAC National Accelerator Laboratory, is supported by the U.S. Department of Energy, Office of Science, Office of Basic Energy Sciences under Contract No. DE-AC02-76SF00515.

**Author contributions**

L.X.Y. and G.H.C. conceived the experiments. Y.D.L. carried out ARPES measurements with the assistance of X.D., W.X.Z, K.Y.Z., Y.Q.H., S.Y.Z., H.K.C, J.Y.L., C.P., Y.H.Y., D.H.L., M.H., Z.K.L., and Y.L.C. *Ab-initio* calculations were performed by X.D. and S.Q.W. DFT+DMFT calculations were performed by Y.L.W. Single crystals were synthesized and characterized by Y.L. and G.H.C. The paper was written by Y.D.L. and L.X.Y. All authors contributed to the scientific planning and discussion.

**Competing interests**

Authors declare that they have no competing interests.

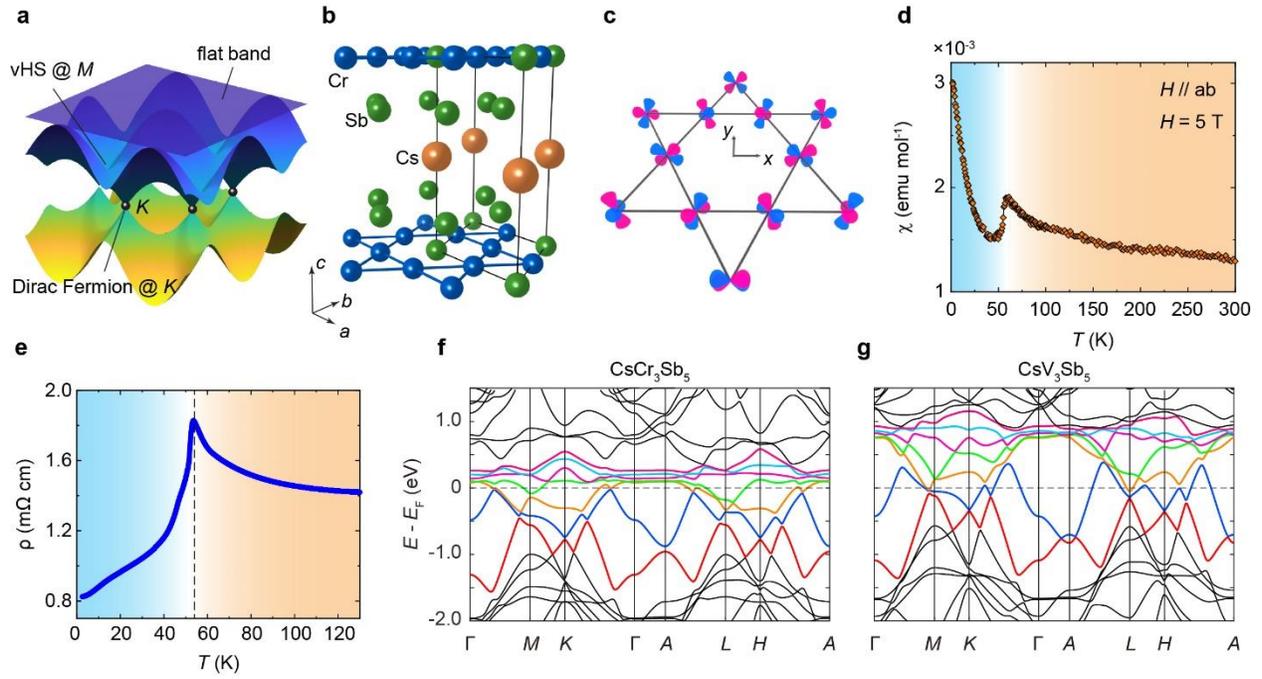

**Fig.1 | Basic properties of $CsCr_3Sb_5$. a**, Schematic illustration of the characteristic band structure of a kagome lattice including the flat band, the Dirac fermion at the *K* point, and the van Hove singularity (vHS) at the *M* point. **b**, The crystal structure of $CsCr_3Sb_5$ in which Cr atoms form planer kagome lattice. **c**, Kagome structure of Cr *d* orbitals. For simplicity, only the $d_{xz}$ orbital is shown. **d**, In-plane magnetic susceptibility as a function of temperature measured at *H* = 5T. **e**, In-plane resistivity as a function of temperature shows a peak near 55 K. **f, g**, Comparison between the calculated band structure of $CsCr_3Sb_5$ (**f**) and $CsV_3Sb_5$ (**g**). The colored lines indicate similar dispersions of the two materials.

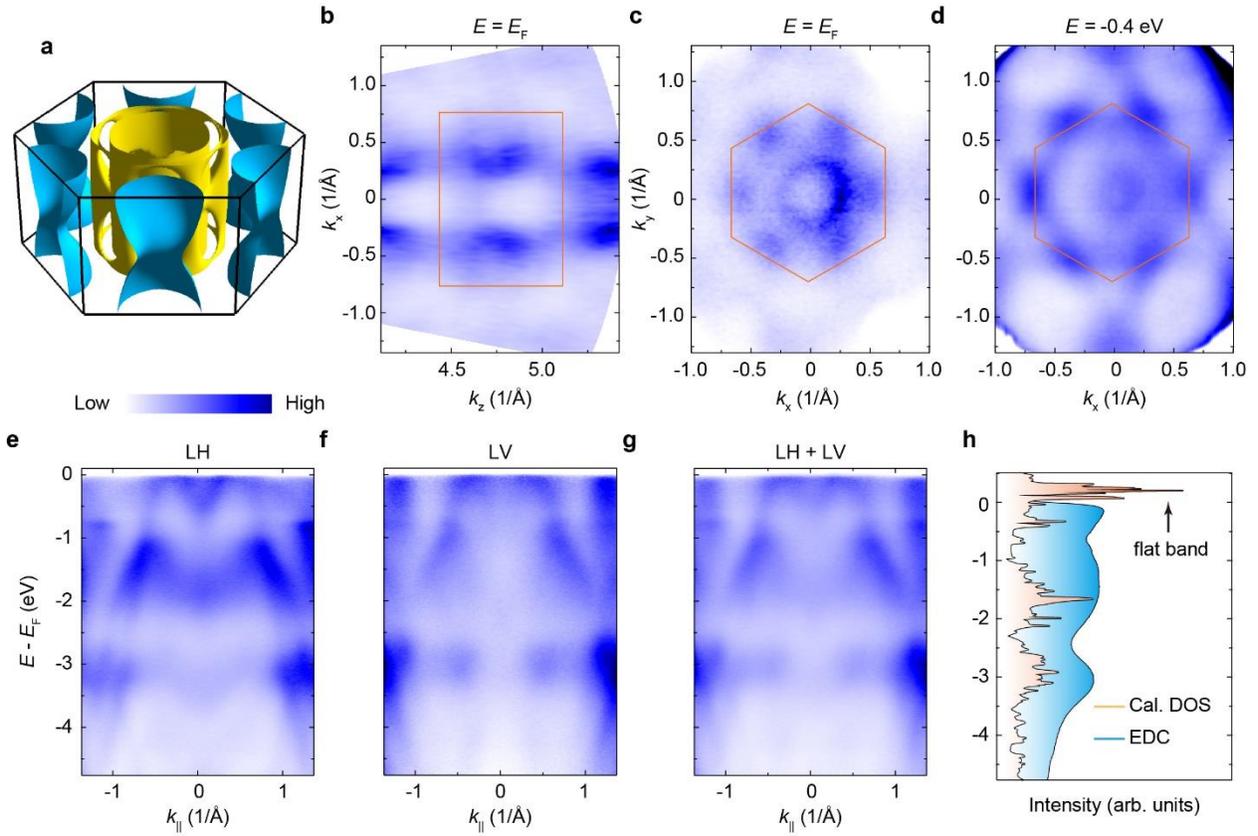

**Fig.2 | Overall band structures of CsCr$_3$Sb$_5$. a**, Three-dimensional plot of the calculated Fermi surface of CsCr$_3$Sb$_5$. **b**, $k_z$ dispersion measured in the photon energy range between 60 eV and 110 eV. **c, d**, The $k_x$-$k_y$ map of ARPES intensity at the Fermi level ($E_F$) and -0.4 eV. **e-g**, Band dispersion along $\bar{\Gamma}\bar{K}$ measured using different photon polarizations in a large energy range. **h**, Comparison between the calculated density of states (DOS) and integrated energy distribution curve (EDC). In panels **c, d,** and **g**, the data measured with linear-horizontally (LH) and linear-vertically (LV) polarized photons were merged for completeness of the electronic structure. Data in **c-g** were measured using 93 eV photons. All data were collected at 6 K.

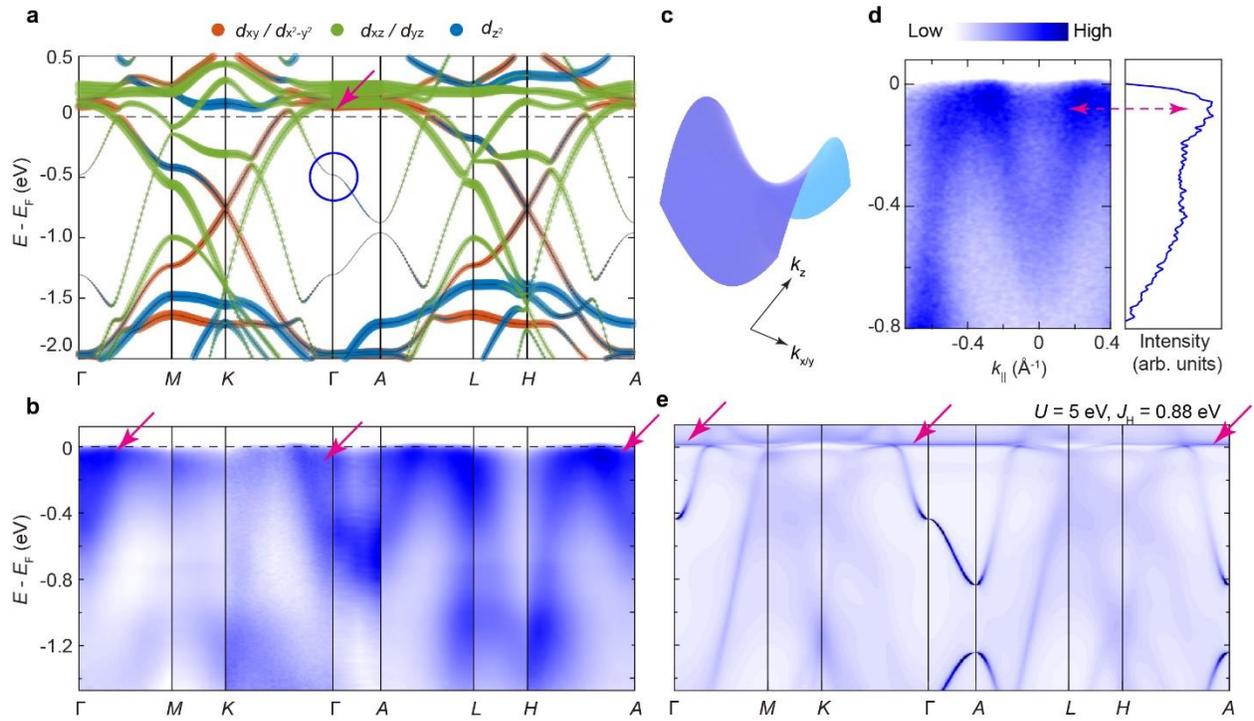

**Fig.3 | Comparison between experimental and calculated fine electronic structure of $CsCr_3Sb_5$ near $E_F$. a**, *Ab-initio* calculation of the orbital-projected electronic structure of $CsCr_3Sb_5$ along high-symmetry directions. **b**, ARPES spectra along high-symmetry directions. **c**, Schematic illustration of the vHS at Γ (blue circle in **a**) with hole-like and electron-like dispersion along $k_z$ and $k_{x/y}$, respectively. **d**, Zoom-in plot of band dispersions along *AH* (left) and the corresponding EDC at *A* (right). **e**, DFT+DMFT calculation of the electronic structure in the paramagnetic state, with on-site Coulomb interaction $U = 5$ eV and Hund's coupling $J_H = 0.88$ eV. The red arrows indicate the flat band close to $E_F$. Data were collected at 6 K. For completeness, data measured with LH- and LV-polarized photons were merged.

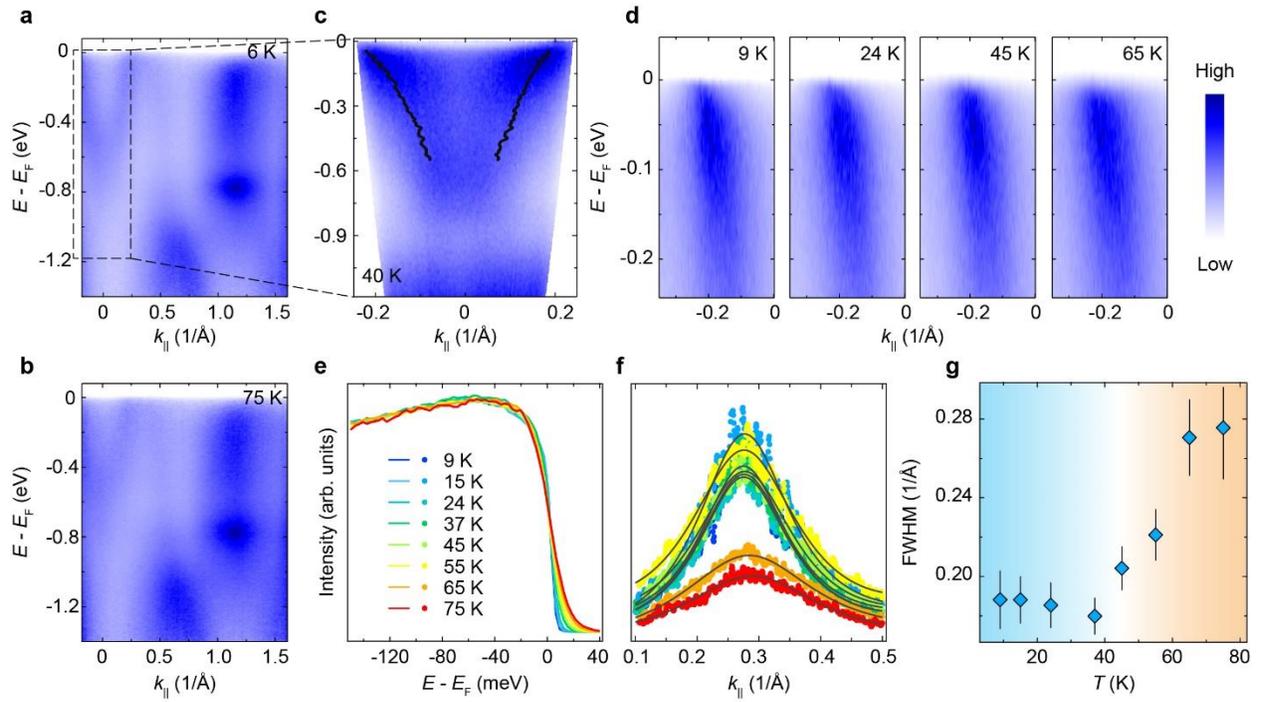

**Fig.4 | Temperature evolution of the electronic structure. a**, **b**, ARPES spectra measured at 6 K (**a**) and 75 K (**b**) with 52 eV photons. **c**, Laser-ARPES spectra measured at 40 K showing the electron band around $\bar{\Gamma}$. The black lines are the extracted band dispersion by fitting the momentum-distribution curves (MDCs) to Lorentzians. **d**, Laser-ARPES spectra measured at different temperatures. **e**, EDCs at the Fermi momentum at different temperatures. **f**, the MDCs at $E_F$ at different temperatures with the fitting results overlaid. The corresponding temperatures are indicated in **e** (dots). **g**, Temperature dependence of the full-width-at-half-maximum (FWHM) of the MDCs at $E_F$. Data in **c-g** were collected using a 7-eV laser.